\documentclass[conference]{IEEEtran}

\IEEEoverridecommandlockouts

\usepackage[letterpaper, top=0.75in, bottom=1.05in, left=0.625in, right=0.625in]{geometry}

\usepackage{cite}
\usepackage{amsmath,amssymb,amsfonts}
\usepackage{algorithmic}
\usepackage{graphicx}
\usepackage{textcomp}
\usepackage{xcolor}
\usepackage{enumitem}
\usepackage{algorithm,algorithmic}
\usepackage{amsthm}
\usepackage[normalem]{ulem}
\usepackage[dvipsnames]{xcolor}  

\def\BibTeX{{\rm B\kern-.05em{\sc i\kern-.025em b}\kern-.08em
 T\kern-.1667em\lower.7ex\hbox{E}\kern-.125emX}}

\IEEEaftertitletext{\vspace{-2em}}
\begin{document}

\title{Two-Dimensional XOR-Based Secret Sharing for Layered Multipath Communication
\thanks{This work was supported in part by National Science Foundation (NSF) under grants CNS2451268, CNS2514415, ONR under grant N000142112472; and the NSF and Office of the Under Secretary of Defense (OUSD) – Research and Engineering, Grant ITE2515378, as part of the NSF Convergence Accelerator Track G: Securely Operating Through 5G Infrastructure Program.}
}

\author{\IEEEauthorblockN{Wai Ming Chan}
\IEEEauthorblockA{\textit{School of Electrical, Computer}\\ \textit{and Energy Engineering} \\
\textit{Arizona State University}\\
Tempe, AZ, USA \\
{wai-ming.chan@asu.edu}}
\and
\IEEEauthorblockN{R\'{e}mi Chou}
\IEEEauthorblockA{\textit{Department of Computer}\\\textit{Science and Engineering} \\
\textit{The University of Texas at Arlington}\\
Arlington, TX, USA \\
{remi.chou@uta.edu}}
\and
\IEEEauthorblockN{Taejoon Kim}
\IEEEauthorblockA{\textit{School of Electrical, Computer}\\ \textit{and Energy Engineering} \\
\textit{Arizona State University}\\
Tempe, AZ, USA \\
{taejoonkim@asu.edu}}
}

\maketitle

\begin{abstract}
This paper introduces the first two-dimensional XOR-based secret sharing scheme for layered multipath communication networks. We present a construction that guarantees successful message recovery and perfect privacy when an adversary observes and disrupts any single path at each transmission layer. The scheme achieves information-theoretic security using only bitwise XOR operations with linear $O(|S|)$ complexity, where $|S|$ is the message length. We provide mathematical proofs demonstrating that the scheme maintains unconditional security regardless of computational resources available to adversaries. Unlike encryption-based approaches vulnerable to quantum computing advances, our construction offers provable security suitable for resource-constrained military environments where computational assumptions may fail. 
\end{abstract}


\section{Introduction}
\label{sec:intro}
Multipath communication improves resilience in adversarial military networks where data traverses partially trusted nodes susceptible to jamming, cyber attacks, and physical destruction \cite{qadir2015,mahmoud2013}. Secret sharing fundamentally differs from encryption: while encryption transforms data to hide it from unauthorized parties, secret sharing divides data into shares such that individual shares reveal nothing about the original secret. This distinction is crucial for multipath networks—encryption protects data confidentiality but cannot ensure availability when paths fail, whereas secret sharing provides both confidentiality (through information-theoretic privacy) and availability (through threshold reconstruction). Moreover, secret sharing offers unconditional security independent of adversarial computational power, unlike encryption which relies on computational hardness assumptions vulnerable to quantum attacks \cite{nist2022}.

\begin{figure}[hbp!]
\centering
\includegraphics[width=0.48\textwidth]{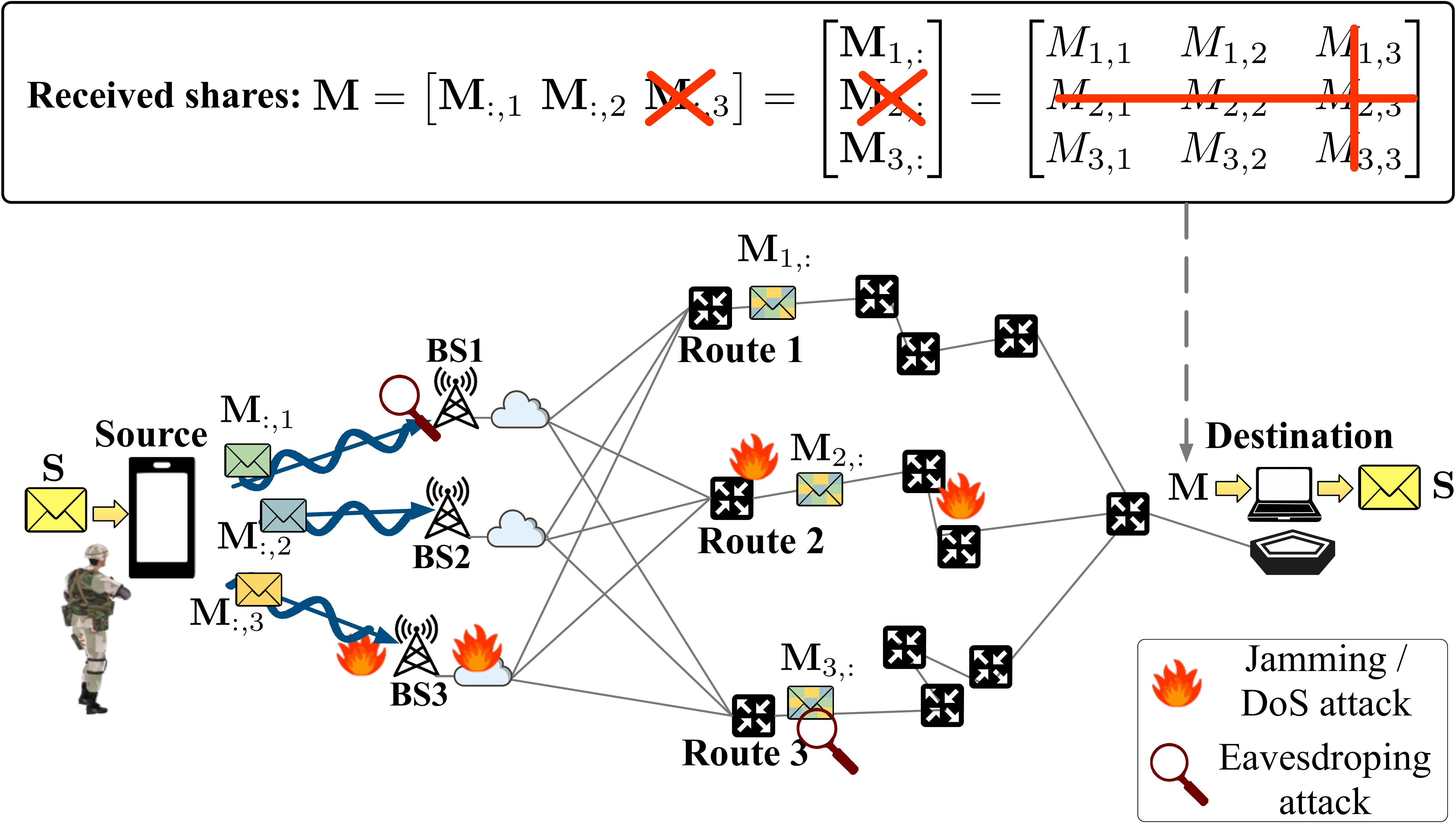}
\caption{Proposed two-layer transmission architecture for secret sharing: The transmitter encodes a secret $S$ into $3\times 3$ matrix, distributes them to 3 BSs, which in turn forward one share each to 3 routes. Each route represents a complete network path that may contain multiple intermediate nodes. The system tolerates the loss or interception of all messages through any single BS and any single route.}
\label{fig:system_two_layer}
\end{figure}

Traditional threshold secret sharing schemes \cite{blakley1979,shamir1979} assume single-hop paths (Fig.~\ref{fig:system_one_layer}) and cannot handle hierarchical failure patterns in layered networks. As shown in Fig.~\ref{fig:system_two_layer}, real networks exhibit correlated cross-layer failures (e.g., jamming at base station 3 (BS $3$) combined with Route $2$ failure) not addressed by existing schemes \cite{shamir1979,lou2004spread, jha2024}.

Polynomial-based schemes like Shamir's \cite{shamir1979} require $O(|S|^2)$ finite field operations for encoding and decoding (where $|S|$ is the message length in bits), making them unsuitable for resource-constrained tactical devices. Encryption-based alternatives cannot prevent DoS attacks and require complex key management infrastructure unsuitable for dynamic military networks. 

The SPREAD scheme \cite{lou2004spread} combines Shamir's secret sharing with node-disjoint routing but assumes single-hop paths. Our prior work \cite{jha2024} introduced XOR-based multipath secret sharing for flat network topologies but does not extend to layered architectures with independent cross-layer failures.

This paper proposes a two-dimensional XOR-based secret sharing scheme for layered multipath networks. The key challenge is maintaining independence between encoding layers while ensuring the secret remains recoverable from any valid subset of shares. Our scheme tolerates any single BS and single route failure while guaranteeing perfect privacy against adversaries observing all transmissions through any single BS and route. Using only XOR operations, our approach achieves information-theoretic security with linear complexity, making it suitable for resource-constrained military devices. 

The remainder of this paper is organized as follows: Section~\ref{sec:system_model} formalizes the system model and design objectives. Section~\ref{sec:one_layer} reviews the one-dimensional baseline construction. Section~\ref{sec:two_dim_scheme} presents our two-dimensional scheme. Section~\ref{sec:security} provides information-theoretic security analysis. Section~\ref{sec:conclusion} provides concluding remarks. Implementation and experimental validation are addressed in complementary systems-focused work, as this paper focuses on establishing the theoretical foundations with mathematical rigor.
\section{System Model and Problem Statement}
\label{sec:system_model}
We consider a secure communication scenario in which a source node transmits a secret message to a destination via a two-layer network architecture shown in Fig.~\ref{fig:system_two_layer}. Throughout this paper, we use ``two-layer'' to refer to the hierarchical network architecture (BSs and routes), while ``two-dimensional'' refers to the resulting $3 \times 3$ matrix structure of shares. The network consists of $ N_1 = 3 $ BSs, each forwarding data along one of $ N_2 = 3 $ routes. We focus on the $3\times 3$ configuration as it represents the minimal non-trivial case that demonstrates the fundamental principles of two-dimensional secret sharing: with $2\times 2$, the scheme degenerates to simple replication, while $3\times 3$ is the smallest configuration requiring careful coordination between eight random sequences to achieve both availability and privacy properties. This configuration serves as a concrete foundation for developing the theoretical framework, with the construction methodology extending naturally to general $N\times M$ topologies. Each route represents a black-box abstraction of a complete network path that may traverse multiple intermediate nodes, routers, and network segments. This abstraction ensures our scheme remains robust to dynamic topology changes within each path, as the internal structure of each black-box route can vary without affecting the encoding scheme. Every communication path between the source and destination is uniquely defined by a BS-route pair, forming a $3 \times 3$ grid of logical transmission paths.

We consider a secret message $S$ that is a binary sequence, where $S \in \{0,1\}^{|S|}$ and $|S|$ denotes the length of $S$ in bits. Before the encoding process, the message $S$ is split into two parts as 
\vspace{-0.25cm}
\begin{equation}
    S = (S_1 \| S_2)
\label{eqn:S_splits}
\end{equation}
where $S_1, S_2 \in \{0,1\}^{|S|/2}$ and $\|$ denotes sequence concatenation. If the original message length is odd, zero-padding is applied for balanced partitioning. 
It is encoded into three intermediate share vectors $ \mathbf{M}_{:,1}, \mathbf{M}_{:,2}, \mathbf{M}_{:,3}$, one for each BS:
\vspace{-0.25cm}
\[
\mathbf{M}_{:,j} = \left[ M_{1,j}, M_{2,j}, M_{3,j} \right]^{T}, \quad \text{for } j = 1,2,3.
\]
Each share vector $ \mathbf{M}_{:,j}$ is routed to BS $j$, which then forwards its components $M_{i,j}$ along routes $i = 1,2,3$.
Route $i$ collects all shares to form a share vector $\mathbf{M}_{i,:} = [M_{i,1}, M_{i,2}, M_{i,3}]^{T}$.
This results in a $3 \times 3$ share matrix $\mathbf{M}$ as
\vspace{-0.1cm}
\begin{equation}
\begin{aligned}
\mathbf{M} = &
\left[
\begin{array}{ccc}
\hspace{0.1cm} M_{1,1} &\hspace{0.3cm} M_{1,2} &\hspace{0.3cm} M_{1,3} \hspace{0.1cm} \\
\hspace{0.1cm} M_{2,1} &\hspace{0.3cm} M_{2,2} &\hspace{0.3cm} M_{2,3} \hspace{0.1cm}\\
\hspace{0.1cm} M_{3,1} &\hspace{0.3cm} M_{3,2} &\hspace{0.3cm} M_{3,3} \hspace{0.1cm}
\end{array}
\right]
\begin{array}{c}
\text{\small{(to route 1)}} \\
\text{\small{(to route 2)}} \\
\text{\small{(to route 3)}}
\end{array}
\\
& \hspace{0.1cm}
\begin{array}{ccc}
\overset{ \text{(to BS 1)} }{} & \hspace{0.2cm}
\overset{ \text{(to BS 2)} }{} & \hspace{0.2cm}
\overset{ \text{(to BS 3)} }{}
\end{array}    
\end{aligned}.
\label{eqn:received_message_matrix}
\end{equation}

\subsection{Information-Theoretic Preliminaries}
\label{sec:info_theory}

We briefly review the information-theoretic concepts used in our security analysis. The \textit{entropy} $H(X)$ of a random variable $X$ measures its uncertainty in bits. For a binary sequence of length $n$, maximum entropy is $n$ bits when all bit patterns are equally likely. The \textit{conditional entropy} $H(X \mid Y)$ quantifies the remaining uncertainty about $X$ after observing $Y$; when $H(X \mid Y) = 0$, $Y$ completely determines $X$. The \textit{mutual information} $I(X;Y) = H(X) - H(X \mid Y)$ measures the information shared between $X$ and $Y$; when $I(X;Y) = 0$, $X$ and $Y$ are independent, meaning knowledge of $Y$ reveals nothing about $X$. This last property is crucial for proving perfect privacy.

\subsection{Adversarial Model and Design Objectives}
\label{sec:design_objectives}

We formalize our threat model considering a computationally unbounded adversary operating in a two-layer network architecture. The adversary has the following capabilities:
\begin{itemize}
\item \textbf{Passive observation:} Can eavesdrop on all transmissions through any single BS (one column of $\mathbf{M}$) and any single route (one row of $\mathbf{M}$) simultaneously.
\item \textbf{Active disruption:} Can jam or disable any single BS and any single route independently through RF interference, cyber attacks, or physical destruction.
\item \textbf{Computational power:} Unbounded computational resources, including potential quantum computing capabilities.
\end{itemize}

Assuming independent failures across layers, we establish two design objectives:

\begin{enumerate}[label=\textbf{O\arabic*}]
  \item \textbf{Availability:}  
    An active adversary may disable one BS and one route independently, for example via RF jamming or denial-of-service (DoS) attacks. This results in the loss of all shares routed through the affected BS (i.e., one column of the share matrix $\mathbf{M}$) and all shares transmitted along the compromised route (i.e., one row). The scheme guarantees recovery of the secret $S$ from the surviving shares. Formally:
{\setlength{\abovedisplayskip}{0.1cm}
 \setlength{\belowdisplayskip}{0.3cm}
\begin{equation}
  \max_{\substack{1 \le r,c \le 3 }}
  H \left(S \mid \left\{ M_{i,j} \mid i \ne r,\; j \ne c \right\} \right) = 0.
  \label{eqn:availability}
\end{equation}
}
  \item \textbf{Perfect Privacy}:
    A passive eavesdropper may observe all shares transmitted through one BS and all shares transmitted along one route, i.e., one full column and one full row of $\mathbf{M}$. The system must ensure that such partial interception does not leak any information about the secret: 
{\setlength{\abovedisplayskip}{0.1cm}
 \setlength{\belowdisplayskip}{0.3cm}
\begin{equation}
  \max_{1 \le r,c \le 3} I \left( \{ M_{i,c} \}_{i=1}^3 \cup \{ M_{r,j} \}_{j=1}^3 ; S \right) = 0.
  \label{eqn:privacy}
\end{equation}
}
\end{enumerate}
\section{Review of Previous Multi-Path Coding Scheme}
\label{sec:one_layer}
\begin{figure}[htbp!]
 \centering
 \includegraphics[width=0.42\textwidth]{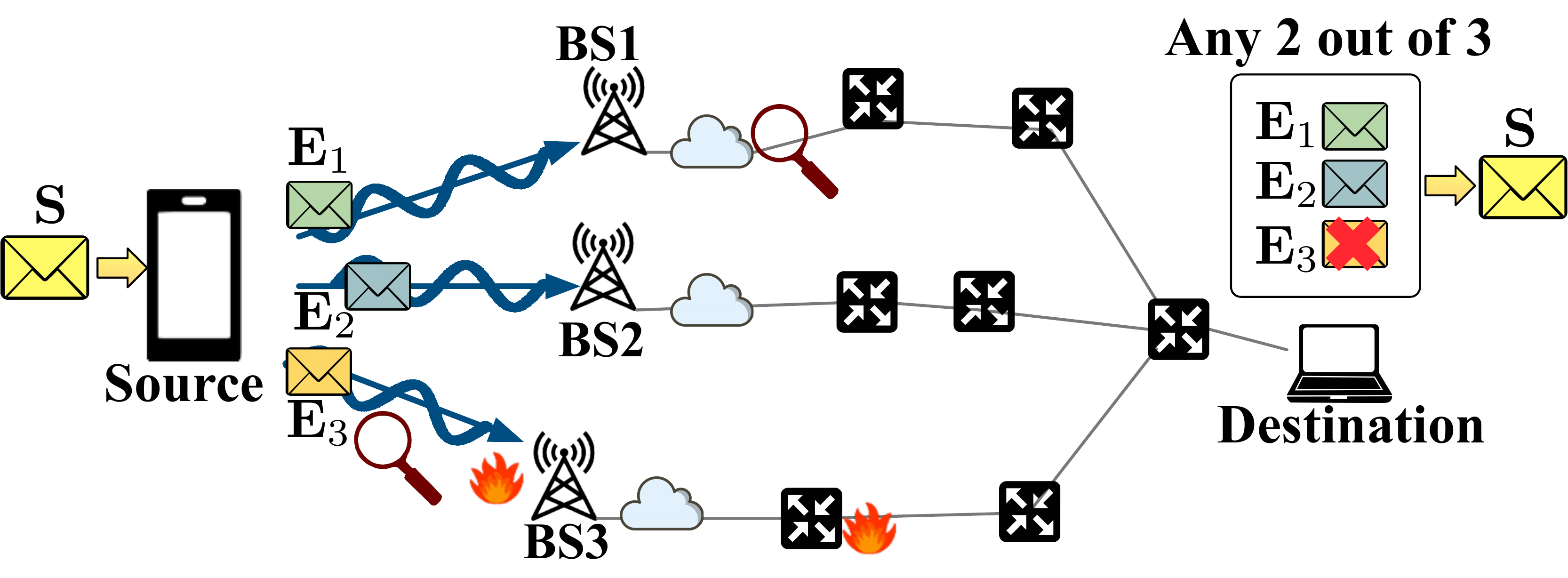}
 \caption{One-layer secret sharing with multi-path transmission. The message is split, encoded into three shares, and sent over separate paths. Any two shares enable recovery; any single share reveals nothing.}
 \label{fig:system_one_layer}
\end{figure}

In \cite{jha2024}, we proposed a secret sharing scheme tailored for multipath communication as shown in Fig. \ref{fig:system_one_layer}. This one-dimensional multi-path coding scheme distributes shares across different communication paths and serves as the foundation for the two-dimensional scheme presented in this paper. The system is designed to tolerate a single path failure, ensuring data availability and perfect privacy, respectively, from dropping or eavesdropping one share.
The approach minimizes computation and storage overhead by utilizing bitwise XOR operations instead of traditional polynomial-based secret sharing methods, making it well-suited for environments with constrained resources and adversarial threats.

\subsection{Encoding Scheme}
\label{sec:encoding_one_layer}
Given a secret message $S$, represented as a binary sequence, the scheme begins by dividing it into two equal-length segments, $S_1$ and $S_2$ in \eqref{eqn:S_splits}.
Two independent binary sequences, $R_1$ and $R_2$ (each matching $S_1$'s length), are generated for encoding the message parts using XOR operations.

The three shares $\mathbf{E}_1$, $\mathbf{E}_2$, and $\mathbf{E}_3$ are generated by
{\setlength{\abovedisplayskip}{0.2cm}
 \setlength{\belowdisplayskip}{0.3cm}
\begin{equation}
 \text{Enc}_{1} (S, R_1, R_2) = \left[\mathbf{E}_1, \mathbf{E}_2, \mathbf{E}_3 \right]^{T} ,
\label{eqn:one_layer_encoding}
\end{equation} 
}
where $\mathbf{E}_1$, $\mathbf{E}_2$, and $\mathbf{E}_3$ are defined as
\vspace{-0.15cm}
\begin{equation}
\begin{aligned}
 \mathbf{E}_1 &= 
 \left( (S_2 \oplus R_2) \| R_1 \right)
 \\
 \mathbf{E}_2 &=
 \left((S_1 \oplus R_1) \| R_2 \right)
 \\
 \mathbf{E}_3 &=
 \left(( S_1 \oplus R_2) \| (S_2 \oplus R_1 ) \right),
\end{aligned}
\label{eqn:one_layer_shares}
\end{equation}
where $\oplus$ denotes the bitwise XOR.
Then, $\mathbf{E}_1, \mathbf{E}_2, \mathbf{E}_3$ are transmitted over three independent communication paths as shown in Fig. \ref{fig:system_one_layer}.

\subsection{Decoding Procedure}
\label{sec:decoding_one_layer}
At the receiver, the original secret $S$ can be reconstructed using any two of the three received shares as illustrated in the Table \ref{tab:one_layer_decoding}. For clarity, we denote each equal-split sequence as $\mathbf{E}_j = \left( \mathbf{E}_j [1] \ \| \ \mathbf{E}_j [2] \right)$ for $j=1,2,3$. 

\vspace{-0.3cm}
\begin{table}[h]
\centering
\caption{Decoding Procedure from Any Two Shares (One-layer)}
\begin{tabular}{|c|p{6.8cm}|}
\hline
\textbf{Case} & \textbf{Reconstructed Secrets} \\
\hline
$\mathbf{E}_1$, $\mathbf{E}_2$ &
Recover $\begin{cases}
 S_1 = \mathbf{E}_1[2] \oplus \mathbf{E}_2[1] \\ S_2 = \mathbf{E}_1[1] \oplus \mathbf{E}_2[2]
\end{cases}$\\
\hline
$\mathbf{E}_1$, $\mathbf{E}_3$ &
Recover $\begin{cases}
 S_1 = \mathbf{E}_1[2] \oplus \mathbf{E}_3[2] \oplus \mathbf{E}_1[1] \oplus \mathbf{E}_3[1] \\ S_2 = \mathbf{E}_1[2] \oplus \mathbf{E}_3[2]
\end{cases}$\\
\hline
$\mathbf{E}_2$, $\mathbf{E}_3$ &
Recover $\begin{cases}
 S_1 = \mathbf{E}_2[2] \oplus \mathbf{E}_3[1] \\ S_2 = \mathbf{E}_2[2] \oplus \mathbf{E}_3[1] \oplus \mathbf{E}_2[1] \oplus \mathbf{E}_3[2]
\end{cases}$ \\
\hline
\end{tabular}
\label{tab:one_layer_decoding}
\end{table}

\subsection{Security Guarantees}
The encoding strategy in \eqref{eqn:one_layer_encoding} offers perfect privacy in the information-theoretic sense. Any single share (e.g., $\mathbf{E}_1$, $\mathbf{E}_2$, or $\mathbf{E}_3$) provides no information about the original secret $S$ \cite{chou2020}, i.e.,
\vspace{-0.15cm}
\begin{equation}
I( \mathbf{E}_j ; S) = 0, \quad \forall j \in \{1,2,3\}.
\label{eqn:privacy_one_layer}
\end{equation}

\section{Proposed Two-Dimensional Secret Sharing Scheme}
\label{sec:two_dim_scheme}

We now present our main contribution: a two-dimensional XOR-based secret sharing scheme that extends the single-layer construction from Section~\ref{sec:one_layer} to a layered network topology described in Section~\ref{sec:system_model}. The scheme distributes shares across a 2D matrix structure to provide availability (O1) and perfect privacy (O2) guarantees against the adversarial model defined in Section~\ref{sec:design_objectives}.

\subsection{Encoding Scheme}

The secret message $S \in \{0,1\}^{|S|}$ is first divided into two equal-length segments $S_1$ and $S_2$ in \eqref{eqn:S_splits}.
The encoding process generates eight independent random binary sequences $R_1, R_2, \ldots, R_8$, each matching $S_1$'s length. These random sequences are generated and known only at the encoder; they are not shared with the decoder.

The message is encoded into a $3 \times 3$ share matrix $\mathbf{M}$, as defined in \eqref{eqn:received_message_matrix}, using a two-layer encoding function $\text{Enc}_2(\cdot)$. This function builds upon the one-layer encoder $\text{Enc}_1(\cdot)$ in \eqref{eqn:one_layer_encoding} and applies it through two sequential rounds of encoding.

First, the message $S$ is encoded using $\text{Enc}_1(\cdot)$ with random sequences $R_1, R_2$ to produce three intermediate shares $(\mathbf{E}_1, \mathbf{E}_2, \mathbf{E}_3) = \text{Enc}_1(S, R_1, R_2)$ in \eqref{eqn:one_layer_shares}.
Then, each intermediate share $\mathbf{E}_j$ is further encoded:
\vspace{-0.1cm}
\begin{equation}
\begin{aligned}
 \text{Enc}_2 \left(S, \{R_j\}_{j=1}^{8} \right) 
 & = \! \big[ 
 \text{Enc}_1 \! \left( \mathbf{E}_1, R_3, R_4 \right), \text{Enc}_1 \! \left( \mathbf{E}_2, R_5, R_6 \right),\\
 & \qquad   \text{Enc}_1 \! \left( \mathbf{E}_3, R_7, R_8 \right) \big] \\
 &= \! \begin{bmatrix}
     \mathbf{M}_{:,1} & \mathbf{M}_{:,2} & \mathbf{M}_{:,3}
 \end{bmatrix}.
\end{aligned}
\label{eqn:two_layer_encoding}
\end{equation}
This yields the following share vectors:
\begin{equation}
\begin{aligned}
 \mathbf{M}_{:,1} & \!=\!\! 
 \begin{bmatrix}
  \left( (R_1 \oplus R_4) \| R_3 \right) \\ 
  \left( (S_2 \oplus R_2 \oplus R_3) \| R_4 \right) \\
  \left( (S_2 \oplus R_2 \oplus R_4) \| (R_1 \oplus R_3) \right)
 \end{bmatrix} \\
 \mathbf{M}_{:,2} & \!=\!\! 
 \begin{bmatrix}
    \left( (R_2 \oplus R_6) \| R_5 \right) \\
    \left( (S_1 \oplus R_1 \oplus R_5) \| R_6 \right) \\
    \left( (S_1 \oplus R_1 \oplus R_6) \| (R_2 \oplus R_5) \right)
  \end{bmatrix} \\
 \mathbf{M}_{:,3} & \!=\!\! 
 \begin{bmatrix}
  \left( (S_2 \oplus R_1 \oplus R_8) \| R_7 \right) \\
  \left( (S_1 \oplus R_2 \oplus R_7) \| R_8 \right) \\
  \left( (S_1 \oplus R_2 \oplus R_8) \| (S_2 \oplus R_1 \oplus R_7) \right)
 \end{bmatrix},
\end{aligned}
\label{eqn:two_layer_shares}
\end{equation}
where $R_3, R_4$ are the additional random sequences for $\mathbf{E}_1$, $R_5, R_6$ for $\mathbf{E}_2$, and $R_7, R_8$ for $\mathbf{E}_3$.

\subsection{Decoding Procedure}
\label{sec:decoding_two_layer}
The two-dimensional scheme can tolerate the failure of one entire row and one entire column of the share matrix $\mathbf{M}$. 
The decoding set is denoted as
\vspace{-0.15cm}
\begin{equation}
\begin{aligned}
 \mathcal{D}_{r,c} &= \{M_{i,j}\mid i\neq r,\;j\neq c\}
\end{aligned}
\label{eqn:submatrix}
\end{equation}
where the entries of $\mathbf{M}$ in row $r$ and column $c$ are missing.

\vspace{-0.1cm}
\begin{algorithm}[hbp]
\caption{Decoding Procedure for 2D Secret Sharing}
\label{alg:two_layer_decoding}
\begin{algorithmic}[1]
\STATE \textbf{Input:} Decoding set $\mathcal{D}_{r,c} = \{M_{i,j} \mid i \neq r, j \neq c\}$ 
\STATE $\mathcal{R} \gets \{1, 2, 3\} \setminus \{r\}$ \hfill \textit{// Available row indices}
\STATE $\mathcal{C} \gets \{1, 2, 3\} \setminus \{c\}$ \hfill \textit{// Available column indices}
\FOR{$j \in \mathcal{C}$}
 \STATE $\mathbf{E}_j \gets \text{Dec}_1 (\{M_{i,j} : i \in \mathcal{R}\} \cap \mathcal{D}_{r,c})$ \hfill \textit{// Recover intermediate share}
\ENDFOR
\STATE $\widehat{S} \gets \text{Dec}_1(\{\mathbf{E}_j : j \in \mathcal{C}\})$ \hfill \textit{// Reconstruct secret}
\RETURN Reconstructed secret $\widehat{S} = S$
\end{algorithmic}
\end{algorithm}

Algorithm~\ref{alg:two_layer_decoding} describes the decoding procedure for all nine failure scenarios, i.e., $1 \leq r,c \leq 3$.
The decoding operations $\text{Dec}_1(\cdot)$ in Steps 5 and 7 follow the one-layer decoding procedures described in Table~\ref{tab:one_layer_decoding}, using any two available shares to reconstruct the target message or intermediate share.

\section{Security Analysis}
\label{sec:security}

We now prove that the proposed two-dimensional scheme satisfies the availability and perfect-privacy objectives stated in \eqref{eqn:availability} and \eqref{eqn:privacy}.

\begin{proof}[Proof of Availability \eqref{eqn:availability}]
We prove that $H(S \mid \mathcal{D}_{r,c}) = 0$ for any $r,c \in \{1,2,3\}$, where $\mathcal{D}_{r,c} = \{ M_{i,j} \mid i \ne r,\; j \ne c \}$.

For each surviving column $j \in \mathcal{C} \triangleq \{1,2,3\} \setminus \{c\}$, the decoding set $\mathcal{D}_{r,c}$ contains exactly two shares from column $j$. 
As demonstrated in Table~\ref{tab:one_layer_decoding}, any two shares from the same column are sufficient to reconstruct the corresponding intermediate message through XOR operations. Therefore, we have
\vspace{-0.25cm}
\begin{equation}
H(\mathbf{E}_j \mid \{M_{i,j} : i \ne r\}) = 0, \quad \forall j \in \mathcal{C}.
\label{eqn:intermediate_recovery}  
\end{equation}

Since $|\mathcal{C}| = 2$, we can recover exactly two intermediate shares. Letting $\{j_1, j_2\} = \mathcal{C}$, we obtain
\vspace{-0.12cm}
\begin{align}
H(\mathbf{E}_{j_1}, \mathbf{E}_{j_2} \mid \mathcal{D}_{r,c}) = 0.
\label{eqn:two_intermediates}
\end{align}

As demonstrated in Table~\ref{tab:one_layer_decoding}, any two intermediate shares from the $\{ \mathbf{E}_1, \mathbf{E}_2, \mathbf{E}_3 \}$ uniquely reconstruct the secret $S$. Therefore, we have
\vspace{-0.2cm}
\begin{equation}
H(S \mid \mathbf{E}_{j_1}, \mathbf{E}_{j_2}) = 0.
\label{eqn:secret_recovery}    
\end{equation}

Since $(\mathbf{E}_{j_1}, \mathbf{E}_{j_2})$ is a deterministic function of $\mathcal{D}_{r,c}$ (by \eqref{eqn:two_intermediates}) and $S$ is a deterministic function of $(\mathbf{E}{j_1}, \mathbf{E}{j_2})$ (by \eqref{eqn:secret_recovery}), the information flow $\mathcal{D}_{r,c} \to (\mathbf{E}_{j_1}, \mathbf{E}_{j_2}) \to S$ forms a Markov chain.
Applying the data processing inequality:
\vspace{-0.1cm}
\begin{align}
H(S \mid \mathcal{D}_{r,c}) 
\leq H(S \mid \mathbf{E}_{j_1}, \mathbf{E}_{j_2}) \nonumber 
= 0.
\label{eqn:final_entropy}
\end{align}
Since this holds for all $r,c \in \{1,2,3\}$, we proved \eqref{eqn:availability}.
\end{proof}

\vspace{0.01cm}
\begin{proof}[Proof of Perfect Privacy \eqref{eqn:privacy}]
Intuitively, any row-column observation reveals only XOR combinations of $S$ with independent random sequences, masking the secret completely.

Formally, we prove that $I(\mathcal{X}_{r,c}; S) = 0$ for any $r,c \in \{1,2,3\}$, where $\mathcal{X}_{r,c} = \{ M_{i,c} \}_{i=1}^3 \cup \{ M_{r,j} \}_{j=1}^3$ denotes the set of shares observed by an adversary who intercepts all transmissions through base station $c$ and all receptions at route $r$.

Since $R_j$ are random sequences independent of the secret $S$, we have 
\vspace{-0.5cm}
\begin{equation}
    I ( \{ R_1, R_2, \ldots, R_8 \} ; S ) = 0 .
\label{eqn:mutual_info_random_key}
\end{equation}

Since each random sequence $R_j$ has the same length as $S_1$ and is uniformly distributed, the entropy of each random sequence is
\vspace{-0.5cm}
\begin{equation}
H (R_j) = \frac{1}{2} |S| , \quad \forall j\in \{1,2,\ldots,8 \}.
\label{eqn:entropy_r}
\end{equation}

We analyze the nine cases for shares observed by the adversary when row $r$ and column $c$ are compromised. In each case, after expanding the observed shares and eliminating terms independent of $S$, we obtain simplified expressions containing only XOR combinations involving $S_1$ or $S_2$:
{\setlength{\abovedisplayskip}{0.1cm}
 \setlength{\belowdisplayskip}{0.08cm}
\begin{align*}
& I(\mathcal{X}_{1,1}; S) = I(S_2 \oplus R_2, R_2 \oplus R_6, S_2 \oplus R_8; S) \\
& I(\mathcal{X}_{2,1}; S) = I(S_1 \oplus R_5, S_1 \oplus R_2 \oplus R_7, S_2 \oplus R_2; S) \\
& I(\mathcal{X}_{3,1}; S) = I (  S_1 \oplus R_6 , S_1 \oplus R_2 \oplus R_8 , S_2 \oplus R_2 , \\
    & \hspace{1.7cm} \qquad  S_2 \oplus R_7 ; S ) 
\\
& I(\mathcal{X}_{1,2}; S) = I(S_1 \oplus R_1, S_2 \oplus R_1 \oplus R_8; S) \\
& I(\mathcal{X}_{2,2}; S) = I(S_1 \oplus R_1, S_1 \oplus R_7, S_2 \oplus R_3; S) \\
& I(\mathcal{X}_{3,2}; S) = I(S_1 \oplus R_1, S_1 \oplus R_8, S_2 \oplus R_4,  \\
    & \hspace{1.7cm} \qquad S_2 \oplus R_1 \oplus R_7; S) \\
& I(\mathcal{X}_{1,3}; S) = I(S_1 \oplus R_2, S_2 \oplus R_1; S) \\
& I(\mathcal{X}_{2,3}; S) = I ( S_1 \oplus R_2, S_2 \oplus R_1, S_2 \oplus R_2 \oplus R_3 ,  \\
    & \hspace{1.7cm} \qquad S_1 \oplus R_1 \oplus R_5 ; S ) \\
& I(\mathcal{X}_{3,3}; S) = I ( S_1 \oplus R_2 , S_2 \oplus R_1 , S_2 \oplus R_2 \oplus R_4 ,  \\
    & \hspace{1.7cm} \qquad  R_1 \oplus R_3 , S_1 \oplus R_1 \oplus R_6 , R_2 \oplus R_5 ; S)  .
\end{align*}
}

We now prove that each mutual information equals zero.

\textbf{Case 1} ($r=1, c=1$):
{\setlength{\abovedisplayskip}{0.1cm}
 \setlength{\belowdisplayskip}{0.08cm}
\begin{align*}
  \!\!\!\!\!\!\!\!
  I(\mathcal{X}_{1,1}; S) 
  &\overset{(a)}{=} H (S_2 \oplus R_2 , R_2 \oplus R_6, S_2 \oplus R_8) \\
  &\quad - H(S_2 \oplus R_2 , R_2 \oplus R_6, S_2 \oplus R_8 \mid S) \\ 
  &\overset{(b)}{\leq}  \frac{3}{2} |S| - H(S_2 \oplus R_2 , R_2 \oplus R_6, S_2 \oplus R_8| S) \\
  &= \frac{3}{2} |S| - H(  R_2 ,   R_6,   R_8 \mid S) \\
  &\overset{(c)}{=} \frac{3}{2} |S| - H(  R_2 ,   R_6,   R_8)  \displaybreak[0]  \\
  &= 0 .  \displaybreak[0] 
\end{align*}
}

\textbf{Case 2} ($r=2, c=1$): 
{\setlength{\abovedisplayskip}{0.1cm}
 \setlength{\belowdisplayskip}{0.08cm}
\begin{align*}
    I(\mathcal{X}_{2,1}; S) 
    &\overset{(a)}{=} H ( S_1 \oplus R_5 , S_1 \oplus R_2 \oplus R_7 , S_2 \oplus R_2 )  \displaybreak[0] \\
    &\quad - H (S_1 \oplus R_5 , S_1 \oplus R_2 \oplus R_7 , S_2 \oplus R_2 \mid S ) \displaybreak[0] \\
    & \overset{(b)}{\leq} \frac{3}{2} |S| \!-\! H (S_1 \oplus R_5 , S_1 \oplus R_2 \oplus R_7 , S_2 \oplus R_2 \! \mid \! S ) \\
    &= \frac{3}{2} |S| - H ( R_5, R_2, R_7, R_2 \mid S)  \\
    &\overset{(c)}{=} \frac{3}{2} |S| - H ( R_5, R_2, R_7 )   \\
  &= 0 .  
\end{align*}
}

\textbf{Case 3} ($r=3, c=1$):
{\setlength{\abovedisplayskip}{0.1cm}
 \setlength{\belowdisplayskip}{0.08cm}
\begin{align*}
    I(\mathcal{X}_{3,1}; S) 
    &\overset{(a)}{=} H \left( \begin{array}{l}
         S_1 \oplus R_6 , S_1 \oplus R_2 \oplus R_8 , \\
    S_2 \oplus R_2 , S_2 \oplus R_7 
    \end{array} \right) \displaybreak[0] \\
    & \quad - H \left( \begin{array}{l}
         S_1 \oplus R_6 , S_1 \oplus R_2 \oplus R_8 , \\
     S_2 \oplus R_2 , S_2 \oplus R_7 
    \end{array} \Bigg| S \right)  \displaybreak[0] \displaybreak[0] \\
    &\overset{(b)}{\leq} 2 |S| - H \left( \!\!\! \begin{array}{l}
         S_1 \oplus R_6 , S_1 \oplus R_2 \oplus R_8 , \\
     S_2 \oplus R_2 , S_2 \oplus R_7 
    \end{array} \Bigg| S \right)  \displaybreak[0] \\
    &= 2 |S| - H (R_6, R_8, R_2, R_7 \mid S) \\
    &\overset{(c)}{=} 2 |S| - H (R_6, R_8, R_2, R_7)   \displaybreak[0] \\
    &= 0 .
\end{align*}
}

\textbf{Case 4} ($r=1, c=2$): 
{\setlength{\abovedisplayskip}{0.1cm}
 \setlength{\belowdisplayskip}{0.08cm}
\begin{align*}
    \!\!\!\!\!\!\!\!\!
    I(\mathcal{X}_{1,2}; S) 
    &\overset{(a)}{=} H (S_1 \oplus R_1, S_2 \oplus R_1 \oplus R_8 ) \\
    & \quad - H ( S_1 \oplus R_1, S_2 \oplus R_1 \oplus R_8 \mid S ) \\
    &\overset{(b)}{\leq} |S| - H ( S_1 \oplus R_1, S_2 \oplus R_1 \oplus R_8 \mid S ) \\
    &= |S| - H ( R_1, R_8 \mid S)  \displaybreak[0] \\
    &\overset{(c)}{=} |S| - H ( R_1, R_8 ) \\
    &= 0 .  \displaybreak[0] 
\end{align*}
}

\textbf{Case 5} ($r=2, c=2$): 
{\setlength{\abovedisplayskip}{0.1cm}
 \setlength{\belowdisplayskip}{0.08cm}
\begin{align*}
    I(\mathcal{X}_{2,2}; S) 
    &\overset{(a)}{=} H ( S_1 \oplus R_1 ,  S_1 \oplus R_7 , S_2 \oplus R_3 ) \\
    & \quad - H ( S_1 \oplus R_1 ,  S_1 \oplus R_7 , S_2 \oplus R_3 \mid S )  \\
    &\overset{(b)}{\leq} \frac{3}{2} |S| - H ( S_1 \oplus R_1 ,  S_1 \oplus R_7 , S_2 \oplus R_3 \mid S )   \\
    &= \frac{3}{2} |S| - H ( R_1, R_7, R_3 \mid S)  \\
    &\overset{(c)}{=} \frac{3}{2} |S| - H ( R_1, R_7, R_3 ) \displaybreak[0] \\
    &= 0.
\end{align*}
}

\textbf{Case 6} ($r=3, c=2$): 
{\setlength{\abovedisplayskip}{0.1cm}
 \setlength{\belowdisplayskip}{0.08cm}
 \begin{align*}
    I(\mathcal{X}_{3,2}; S) 
    &\overset{(a)}{=} H \left( \begin{array}{l}
          S_1 \oplus R_1 , S_1 \oplus R_8 , \\
     S_2 \oplus R_4, S_2 \oplus R_1 \oplus R_7 
    \end{array} \right) \\
    & \quad - H \left( \begin{array}{l}
          S_1 \oplus R_1 , S_1 \oplus R_8 , \\
     S_2 \oplus R_4, S_2 \oplus R_1 \oplus R_7 
    \end{array} \Bigg| S \right) \\
    &\overset{(b)}{\leq} 2 |S| - H \left( \begin{array}{l}
          S_1 \oplus R_1 , S_1 \oplus R_8 , \\
     S_2 \oplus R_4, S_2 \oplus R_1 \oplus R_7 
    \end{array} \Bigg| S \right) \\
    &= 2 |S| - H ( R_1, R_8, R_4, R_7 \mid S) \\
    &\overset{(c)}{=} 2 |S| - H ( R_1, R_8, R_4, R_7 ) \\
    &= 0 .
\end{align*}
}

\textbf{Case 7} ($r=1, c=3$): 
{\setlength{\abovedisplayskip}{0.1cm}
 \setlength{\belowdisplayskip}{0.08cm}
\begin{align*}
    I(\mathcal{X}_{1,3}; S) 
    &\overset{(a)}{=} H (S_1 \oplus R_2 , S_2 \oplus R_1) \\
    &\quad - H (S_1 \oplus R_2 , S_2 \oplus R_1 \mid S)   \displaybreak[0]  \\
    &\overset{(b)}{\leq} |S| - H(S_1 \oplus R_2 , S_2 \oplus R_1 \mid S) \\
    &= |S| - H(R_2, R_1 \mid S) \displaybreak[0] \\
    &\overset{(c)}{=} |S| - H(R_2, R_1) \\
    &= 0 .
\end{align*}
}

\textbf{Case 8} ($r=2, c=3$): 
{\setlength{\abovedisplayskip}{0.1cm}
 \setlength{\belowdisplayskip}{0.08cm}
 \begin{align*}
    I(\mathcal{X}_{2,3}; S) 
    &\overset{(a)}{=} H \left( \begin{array}{l}
         S_1 \oplus R_2, S_2 \oplus R_2 \oplus R_3 , \\
         S_2 \oplus R_1, S_1 \oplus R_1 \oplus R_5
    \end{array} \right) \\
    & \quad - H \left( \begin{array}{l}
         S_1 \oplus R_2, S_2 \oplus R_2 \oplus R_3 , \\
         S_2 \oplus R_1, S_1 \oplus R_1 \oplus R_5
    \end{array} \!\! \Bigg| S \right)  \displaybreak[0] \\
    &\overset{(b)}{\leq} 2 |S| - H \left( \!\! \begin{array}{l}
         S_1 \oplus R_2, S_2 \oplus R_2 \oplus R_3 , \\
         S_2 \oplus R_1, S_1 \oplus R_1 \oplus R_5
    \end{array} \!\! \Bigg| S \right)  \displaybreak[0] \\
    &= 2 |S| - H (R_1, R_2, R_3, R_5 \mid S) \\
    &\overset{(c)}{=} 2 |S| - H (R_1, R_2, R_3, R_5 )  \displaybreak[0]  \\
    &= 0 .  \displaybreak[0] 
 \end{align*}
}

\textbf{Case 9} ($r=3, c=3$): 
{\setlength{\abovedisplayskip}{0.1cm}
 \setlength{\belowdisplayskip}{0.08cm}
 \begin{align*}
    I(\mathcal{X}_{3,3}; S) &=
    H \left( \!\! \begin{array}{c}
        S_1 \oplus R_2 , S_2 \oplus R_1 , S_2 \oplus R_2 \oplus R_4 , \\
        R_1 \oplus R_3 , S_1 \oplus R_1 \oplus R_6 , R_2 \oplus R_5
    \end{array} \right)   \displaybreak[0] \\
    &\!\!\quad - H \left( \!\! \begin{array}{c}
          S_1 \oplus R_2 , S_2 \oplus R_1 , S_2 \oplus R_2 \oplus R_4 , \\
        R_1 \oplus R_3 , S_1 \oplus R_1 \oplus R_6 , R_2 \oplus R_5
    \end{array} \!\! \Bigg| S \right)   \\
    &\!\!\overset{(b)}{\leq} 3 |S| \!-\! H \! \left( \!\!\!\! \begin{array}{c}
          S_1 \oplus R_2 , S_2 \oplus R_1 , S_2 \oplus R_2 \oplus R_4 , \\
        R_1 \oplus R_3 , S_1 \oplus R_1 \oplus R_6 , R_2 \oplus R_5
    \end{array} \!\! \Bigg| S \! \right) \\
    &\!\!= 3 |S| - H (R_1, R_2, R_3, R_4, R_5, R_6 \mid S) \\
    &\!\!\overset{(c)}{=}  3 |S| - H (R_1, R_2, R_3, R_4, R_5, R_6) \\
    &\!\!= 0.
\end{align*}
}

In all nine cases, steps (a), (b), and (c) follow the same reasoning: (a) applies the definition of mutual information, (b) uses that entropy is maximized by uniform distribution, and (c) follows from the independence between $S$ and all $R_j$'s. Since $I(\mathcal{X}_{r,c}; S) = 0$ for all $r,c \in \{1,2,3\}$, perfect privacy \eqref{eqn:privacy} is achieved.
\end{proof}

\section{Conclusion}
\label{sec:conclusion}
This paper introduced the first two-dimensional XOR-based secret sharing scheme for layered multipath communication networks, with information-theoretic security proofs. The scheme guarantees availability and perfect privacy when any single base station and any single route fail simultaneously. Our mathematical proofs establish unconditional security using only XOR operations, making the scheme practical for resource-constrained military devices. The $3\times 3$ configuration serves as both a practical expansion of network resilience through three paths in a two-layer structure and the minimal non-trivial case for achieving both availability and perfect privacy. 


\bibliographystyle{IEEEtran}
\bibliography{chan}

\begin{thebibliography}{1}
\providecommand{\url}[1]{#1}
\csname url@samestyle\endcsname
\providecommand{\newblock}{\relax}
\providecommand{\bibinfo}[2]{#2}
\providecommand{\BIBentrySTDinterwordspacing}{\spaceskip=0pt\relax}
\providecommand{\BIBentryALTinterwordstretchfactor}{4}
\providecommand{\BIBentryALTinterwordspacing}{\spaceskip=\fontdimen2\font plus
\BIBentryALTinterwordstretchfactor\fontdimen3\font minus \fontdimen4\font\relax}
\providecommand{\BIBforeignlanguage}[2]{{%
\expandafter\ifx\csname l@#1\endcsname\relax
\typeout{** WARNING: IEEEtran.bst: No hyphenation pattern has been}%
\typeout{** loaded for the language `#1'. Using the pattern for}%
\typeout{** the default language instead.}%
\else
\language=\csname l@#1\endcsname
\fi
#2}}
\providecommand{\BIBdecl}{\relax}
\BIBdecl

\bibitem{qadir2015}
J.~Qadir, A.~Ali, K.-L.~A. Yau, A.~Sathiaseelan, and J.~Crowcroft, ``Exploiting the power of multiplicity: a holistic survey of network-layer multipath,'' \emph{IEEE Communications Surveys \& Tutorials}, vol.~17, no.~4, pp. 2176--2213, 2015.

\bibitem{mahmoud2013}
M.~Z. Hasan, H.~Al-Rizzo, and F.~Al-Turjman, ``A survey on multipath routing protocols for {QoS} assurances in real-time wireless multimedia sensor networks,'' \emph{IEEE Communications Surveys \& Tutorials}, vol.~19, no.~3, pp. 1424--1456, 2017.

\bibitem{nist2022}
G.~Alagic, G.~Alagic, D.~Apon, D.~Cooper, Q.~Dang, T.~Dang, J.~Kelsey, J.~Lichtinger, Y.-K. Liu, C.~Miller \emph{et~al.}, ``Status report on the third round of the {NIST} post-quantum cryptography standardization process,'' National Institute of Standards and Technology, Tech. Rep., 2022.

\bibitem{blakley1979}
G.~R. Blakley, ``Safeguarding cryptographic keys,'' \emph{Proceedings of AFIPS}, vol.~48, pp. 313--317, 1979.

\bibitem{shamir1979}
A.~Shamir, ``How to share a secret,'' \emph{Communications of the ACM}, vol.~22, no.~11, pp. 612--613, 1979.

\bibitem{lou2004spread}
W.~Lou, W.~Liu, and Y.~Fang, ``{SPREAD}: enhancing data confidentiality in mobile ad hoc networks,'' in \emph{IEEE INFOCOM 2004}, vol.~4, 2004, pp. 2404--2413.

\bibitem{jha2024}
A.~Jha, S.~Kashani, M.~Hossein, A.~Kirchner, M.~Zhang, R.~A. Chou, S.~W. Kim, H.~M. Kwon, V.~Marojevic, and T.~Kim, ``Enhancing next{G} wireless security: A lightweight secret sharing scheme with robust integrity check for military communications,'' in \emph{MILCOM 2024 - 2024 IEEE Military Communications Conference (MILCOM)}, 2024, pp. 1--6.

\bibitem{chou2020}
R.~A. Chou and J.~Kliewer, ``Secure distributed storage: Rate-privacy trade-off and {XOR}-based coding scheme,'' in \emph{2020 IEEE International Symposium on Information Theory (ISIT)}, 2020, pp. 605--610.

\end{thebibliography}

\end{document}